# BEAM DIAGNOSTICS AT DAΦNE WITH FAST UNCOOLED IR DETECTORS

A. Bocci, A. Clozza, A. Drago, A. Grilli, A. Marcelli, M. Piccinini, A. Raco, R. Sorchetti,
INFN/LNF, Frascati, Italy
Lisa Gambicorti, INOA, Firenze, Italy
A. De Sio, E. Pace, Università degli Studi di Firenze, Firenze, Italy
J. Piotrowski, Vigo System Sa, Warsaw, Poland

*Abstract*

Bunch-by-bunch longitudinal diagnostics is a key issue of modern accelerators. To face up this challenging demand, tests of mid-IR compact uncooled photo-conductive HgCdTe detectors have been recently performed at DAΦNE. Different devices were used to monitor the emission of e$^-$ bunches. The first experiments allowed recording of 2.7 ns long e$^-$ bunches with a FWHM of a single pulse of about 600 ps. These results address the possibility to improve diagnostics at DAΦNE and to this purpose an exit port on a bending magnet of the positron ring has been set-up. An HV chamber, hosting a gold-coated plane mirror that collects and deflects the radiation through a ZnSe window, is the front-end of this port. After the window, a simple optical layout in air allows focusing IR radiation on different detectors. The instrumentation will allow comparison in the sub-ns time domain between the two rings and to identify and characterize bunch instabilities. Moreover, to improve performances tests of new photovoltaic detectors with sub-ns response times are in progress. We will briefly summarize the actual status of the 3+L experiment and will discuss future applications of fast IR photovoltaic detectors and the development of fast IR array detectors.

## INTRODUCTION

Beam diagnostics is an essential component of a particle accelerator. All storage rings emit synchrotron light and different radiation energies can be used for beam diagnostics. Indeed, the synchrotron radiation emission covers a wide energy range from IR to X-ray energies with a pulsed structure that depends by the temporal characteristic of the stored beam. The radiation can be used to monitor the beam stability and to measure the longitudinal profiles of the accumulated particles. However, due to the time structure of the synchrotron radiation, to perform beam diagnostics fast detectors are required. At third generation synchrotron radiation sources useful devices for the diagnostics of accelerated particles need response times from the ns to the ps range while the future FEL sources need response time in the fs domain.

The main advantage of photon diagnostic is that it is a direct and non-destructive probe. Diagnostic based on synchrotron light is typically used for imaging and allows the measure of the beam cross section as well as the longitudinal structure such as the bunch length of stored particles. The determination of the bunch length is an important operational parameter of storage rings that allow monitoring the beam dynamics.

The main requirements of a beam diagnostic system could be: fast, at least in the sub-ns regime to guarantee the installation in all accelerators, compact and robust. However, it could be also easy to manage, possibly vacuum compatible and at low cost.

Standard beam diagnostic methods use a streak camera, an extremely fast photon detectors that takes an instantaneous image of the particles running along the orbit. Images of the temporal structure of a beam have resolution time of ~1 ps or below [1]. Streak cameras are powerful detectors whose principal drawback is the cost. Moreover, streak cameras are complex and fragile devices and usually are not used for permanent full-time beam diagnostics. As a consequence fast, cheaper and compact photon detectors may represent an important alternative for photon beam diagnostics. Photon devices are also easier to manage with respect to a streak camera. The availability of uncooled infrared devices optimized for the mid-IR range, based on HgCdTe alloy semiconductors, already now allow obtaining sub-ns response times [2]. These detectors can be used for fast detection of the intense synchrotron radiation IR sources and then for beam diagnostics. Preliminary measurements of the pulsed synchrotron light emission have been performed with uncooled IR photo-conductive detectors at DAΦNE, the e$^+$-e$^-$ collider of the LNF laboratory of the Istituto Nazionale di Fisica Nucleare (INFN), achieving a resolution time of about few hundred of picoseconds [3,4]. In this contribution we will present and discuss preliminary results obtained with such photo-conductive detector.

Experiments have been performed at SINBAD (Synchrotron Infrared Beamline At DAΦNE), the IR beamline operational at Frascati since 2001. Moreover, to improve the DAΦNE diagnostics a new experiment, 3+L (*Time Resolved Positron Light Emission*), funded by INFN, started the installation at the exit of one of the bending magnet of the DAΦNE positron ring. The installation, when completed, will allow monitoring the positron bunch lengths at DAΦNE with the main aim to study and characterize the instabilities of the positron beam and in order to possibly increase the positron current and the collider luminosity.

A short description of the 3+L experiment, the optical simulations and the actual status of the experiment will be presented. Finally we will discuss how to improve the time resolution using faster IR photovoltaic detectors and how to perform transverse diagnostics with new IR array detectors working at room temperature.

## MEASUREMENTS OF THE ELECTRON BEAM ON THE SINBAD BEAMLINE

Preliminary measurements of the pulsed signal of the electron beam have been carried out at DAΦNE with uncooled IR detectors. Experiments have been performed using the IR emission collected at the end of the optical system of SINBAD [5]. As showed in Fig. 1 the IR beam is extracted from a bending magnet located in the external arc section of the electron ring. DAΦNE is the Frascati $e^+/e^-$ collider, with a center of mass energy of 1.02 GeV, designed to operate at high current (>2 A) and up to 120 bunches [6]. Different bunch patterns can be stored at DAΦNE but the minimum bunch distance is 2.7 ns with a maximum achieved single-bunch current of ~20 mA. Bunches have a quasi-Gaussian shape with a length in the range of 100 - 300 ps FWHM. For the first experiments we used a single-element uncooled photo-detector trying to resolve the time structure of the IR light emission. The detector has been placed at the focus of the last optical mirror just at the entrance slit of the interferometer.

The emission of 105 bunches and the gap between the last and the first bunch was measured with an average current of ~3 mA per bunch. The photo-conductive detector was biased through a bias tee with a current of about 20 mA. The output signal was amplified by a voltage amplifier with a bandwidth of 2.5 GHz and a gain of ~40 dB. Acquisitions have been performed with a scope (model Tektronix TDS 820) with a bandwidth of 6 GHz. The average response time of the detector was of ~600 ps and the rise time of the signal emitted by a single-electron bunch was ~400 ps. The amplified signal of the IR detector has been used also to perfom measurements of the longitudinal feedback of the electron ring using a spectrum analyzer. In the multibunch mode, the electron synchrotron frequency has been measured as well as the correct longitudinal feedback behaviour [4].

## POSITRON BEAM DIAGNOSTICS: THE 3+L EXPERIMENT

To test and characterize a simple set up based on such uncooled IR detectors and with the objective to perform bunch by bunch beam diagnostics on the positron ring, a compact experimental installation was set up inside the DAΦNE hall in the framework of the 3+L experiment. In Fig. 1 is showed the layout of the DAΦNE complex and the location of the 3+L exit port. The experimental system is placed after the IP2 interaction region collecting the light from a bending magnet having a critical energy of 273 eV, on the only available exit port of the positron ring.

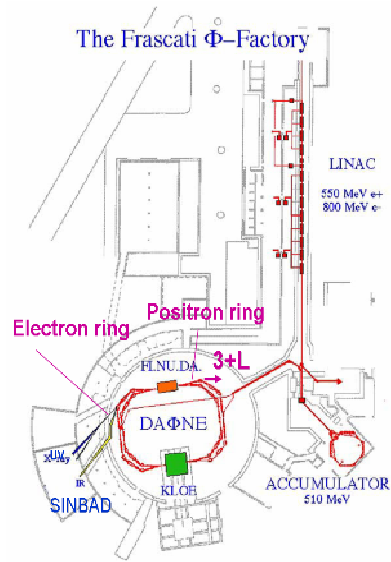

Figure 1 : Layout of the DAΦNE collider. The 3+L exit port inside the hall and of the SINBAD beamline are also showed.

The experiment under installation on this exit port is outlined in Fig. 2. It consists in a simple front-end where a HV chamber hosts a gold-coated plane mirror that collects and deflects the IR radiation through a ZnSe window. This latter window allows transmission of IR radiation between 0.6 to 12 μm. After the window, a simple optical layout composed by 5 mirrors in air will focus the radiation on IR detectors as illustrated in Fig. 2.

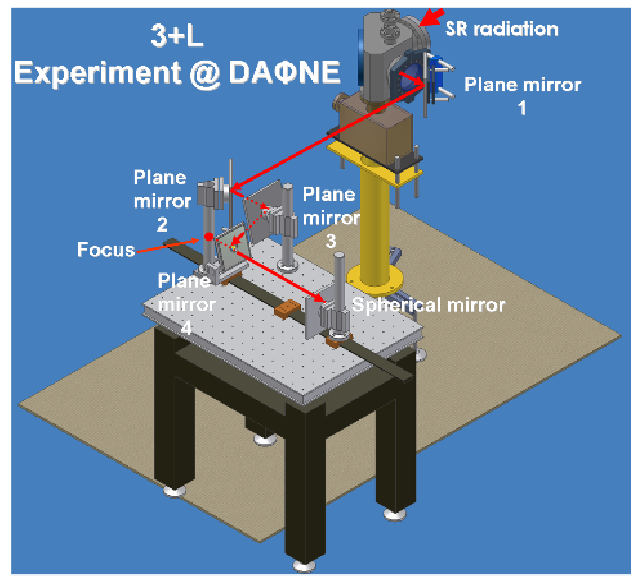

Figure 2 : CAD drawing of the optical system of the 3+L experiment. The IR beam path is outlined by red lines showing the last mirror with the central hole.

Ray tracing simulations have been performed to design and optimize the optical system. The optical layout is based on four plane mirrors that transport 60 x 20 mrad$^2$ of the emitted radiation to the optical table where a spherical mirror will focus the radiation on the IR detector.

Different detectors could be aligned and tested thanks to a motorized and remotely controlled xyz stages. The remotely controlled scope connected to a PC by a GPIB I/O controller will be used to collect data. A dedicated software package for acquisition has been developed under the Labview platform.

To optimize the optical system and to compare the measured intensity of the IR source we performed simulations at the wavelength of 10 µm with the SRW software package [7]. In particular, we calculated the intensity and the distribution of the vertical and horizontal intensity of the IR synchrotron radiation source at the exit port before the first gold coated mirror installed inside the HV chamber.

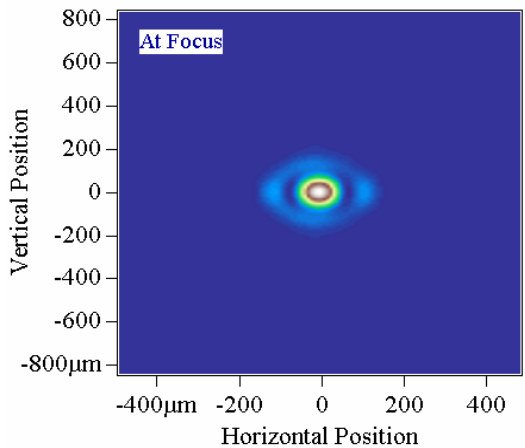

Figure 3 : The spot size at 10 µm simulated at the focus of the optical system outlined in Fig. 2.

To calculate the flux of the source at the focus of the optical system we performed different simulations. To characterize the power of the source at the exit of the window with a calibrated NIST power meter we have also performed preliminary measurements. Data have been collected with the power-meter model Melles Griot 13 PEM 001/J with different IR filters in the range 5-20 µm. The power of the source measured after the first reflection on the first mirror of the optical system is ~0.08 mW. A comparison between the measurements, the simulations and the calculated data is now in progress.

From simulations we obtain that more than 50 % of the energy of the source at the exit port, in the range 0.6-10 µm is contained inside a 400 x 400 micron$^2$ spot at the focus of the optical system. The simulated spot size at 10 µm is showed in Fig. 3.

When the optical system will be installed and aligned a measurement of the power available at the focus of the beam will allow a reliable evaluation of the beam loss of the mirror layout.

When completed, the main goal of the 3+L experimental apparatus will be the systematic acquisition of the positron bunch profiles in order to identify with a real-time method beam instabilities and to improve the existing DAΦNE diagnostics on the positron ring. Positron bunch instabilities have been observed at DAΦNE and actually limit the maximum available current to ~1.3 A. These instabilities have been associated to the occurrence of e$^-$ cloud effect inside the pipe [8,9]. Detailed and simultaneous measurements and comparison of bunch lengths in both electron and positron rings by fast photon detectors could be very helpful to characterize these phenomena, not only to improve the DAΦNE performances but also to shed light on the origin of the instabilities. We will try also to understand the possible role of e$^-$ cloud effects and how it may affect the maximum current in the DAΦNE positron ring. In conclusion, this optical diagnostics approach could be important to increase the accumulated current in the DAΦNE e$^+$ ring and consequently the collider luminosity.

## FUTURE APPLICATIONS

New photo-voltaic IR detectors operating at near room temperature (from 300 K to 205 K temperature) with faster response time will be also test to further improve longitudinal beam diagnostics. The response time of such devices is of the order of 100 ps or lower. Also these devices are based on HgCdTe multilayer heterostructures grown by MOCVD on (211) and (111) GaAs substrates [10,11]. Preliminary characterization of these photo-voltaic devices has been performed at SINBAD with success but recently additional tests were performed and analysis is in progress.

Finally a fast IR array detector has been developed in collaboration with the VIGO SA [3]. The array showed in Fig. 4 is a customized uncooled IR photoconductive matrix constituted by 32x2 pixels. The size of each single pixel is about 50x50 µm and their response time is about 500 ps. A first experimental characterization of the array is planned on the SINBAD beamline after the completion of the electronics that is foreseen before summer.

With such device a bunch by bunch imaging of the IR source and investigations of the transverse bunch instabilities will be also possible.

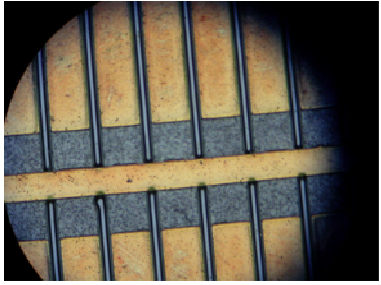
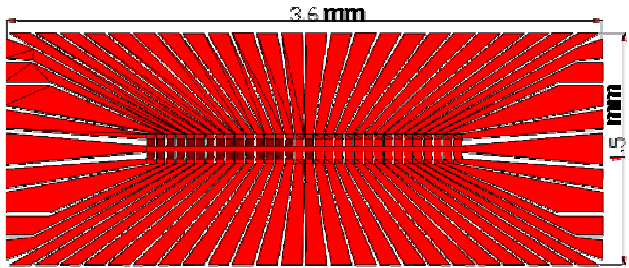

Figure 4 : Top: a small detail of the IR photoconductive array detector. Bottom: layout of the array showing its real dimension.

## CONCLUSIONS

A new method to perform bunch-by-bunch longitudinal and transverse measurements is under test at the DAΦNE collider. To monitor the bunch profile on both rings of this low energy and high current collider this optical method employs fast IR devices. The IR photo-conductive or photo-voltaic devices we use are made by MCT semiconductors and work at near room temperature. They are fast, robust, vacuum compatible, easy to manage and in particular are available at much lower cost if compared with fast alternative devices such as streak-cameras.

Measurements performed at DAΦNE at SINBAD beamline with uncooled photo-conductive detectors looking at the time structure of the electron bunches showed a sub-ns response time. To improve beam diagnostics on the positron ring a novel experiment is under installation at the exit port of one of the bending magnet of the positron ring. The instrumentation will be used to monitor the bunch profiles of the stored positrons, i.e., measurements of bunch-by-bunch lengths and to identify and characterize positron beam instabilities. Data could be useful to increase the current on the $e^+$ ring and the collider luminosity. Future foreseen applications of the technology are based on faster photo-voltaic devices with <100 ps response time and IR uncooled array detectors to achieve bunch by bunch imaging of the source and to investigate simultaneously transverse bunches instabilities on the DAΦNE rings. A first prototype made by 32x2 pixels each 50x50 $\mu m^2$ will be soon characterized on the SINBAD beamline.

## ACKNOWLEDGMENTS

Special thanks are due to the DAΦNE accelerator group for running dedicated beamtime at the 3+L experiment, to M. Pietropaoli and G. Viviani for the technical assistance in the DAFNE-Light laboratory, to M.A. Frani for the support on the software development, to G. Cinque for fruitful discussions and support with the SRW SW package.